          [2001/04/25 1.1 (PWD)]
\documentclass[a4paper,twocolumn]{esapub} % European paper
\usepackage{natbib}
\usepackage{graphicx}

\title{Pixelized Gas Micro-Well Detectors for Advanced Gamma-Ray Telescopes}
\author{Peter F. Bloser}
\author{Stanley D. Hunter}
\affil{NASA/Goddard Space Flight Center, Greenbelt, MD 20771, USA, bloser@gsfc.nasa.gov}

\begin{document}


\maketitle

\begin{abstract}
We describe possible applications of pixelized micro-well detectors (PMWDs)
as three-dimensional charged particle trackers in advanced
gamma-ray telescope concepts.  A micro-well detector consists of an array of
individual micro-patterned gas proportional counters opposite a planar drift
electrode.  When combined with pixelized thin film transistor (TFT) array
readouts, large gas volumes may be imaged with very good spatial and energy
resolution at reasonable cost.  The third dimension is determined by
timing the drift of the ionization electrons.  The primary advantage of
this technique is the very low scattering that the charged particles
experience in a gas tracking volume, and the very accurate determination
of the initial particle momenta that is thus achieved.  We consider two
applications of PMWDs to gamma-ray astronomy: 1) A tracker for an
Advanced Compton Telescope (ACT) in which the recoil electron from the
initial Compton scatter may be accurately tracked, greatly reducing the
telescope's point spread function and increasing its polarization sensitivity;
and 2) an Advanced Pair Telescope (APT) whose angular resolution is limited
primarily by the nuclear recoil and which achieves useful polarization
sensitivity near 100 MeV. We have performed Geant4 simulations of both
these concepts to estimate their angular resolution and sensitivity for
reasonable mission designs.
\end{abstract}

\section{Introduction}

The next generation of medium-energy (0.5 -- 50 MeV) and high-energy (30 MeV -- 100 GeV) 
gamma-ray telescopes (Compton scatter
and pair production telescopes, respectively) will require a substantial improvement
in angular resolution in order to greatly improve on the sensitivity of previous and 
currently-planned
missions.  In both cases, accurate imaging, which decreases the relative influence of 
background, relies on a good knowledge of the momenta of secondary particles produced in 
the primary gamma-ray interaction.  These secondary particles are the scattered gamma-ray
and recoil electron in the case of Compton scattering, and the electron-positron pairs
in the case of pair production.  Precisely recording these momenta also enables various
background-rejection techniques and greatly increases the sensitivity of the telescope to
the polarization of the incident radiation.

The initial secondary particle momenta are masked by poor spatial resolution
and by multiple Coulomb scattering of charged particles within the detector materials.  
These factors have contributed to an enlarged point spread function (PSF) in current
gamma-ray instruments and, in the case of pair production telescopes, have totally 
suppressed the polarization sensitivity.  Improving this picture will require 
a low-density tracking medium with high spatial readout resolution.  We therefore
propose basing future gamma-ray instruments on micro-pattern gas detectors.  Here we
outline possible designs for Compton and pair telescopes using pixelized gas micro-well
detectors under development at NASA/GSFC.

\section{Pixelized Micro-Well Detectors}

\begin{figure}
\centering
\includegraphics[width=5cm]{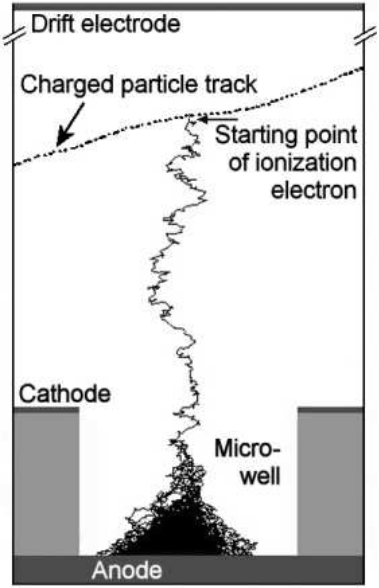}
\caption{Schematic (not to scale) of a micro-well detector.  Charge amplification
takes place in a high electric field inside the well.\label{fig:well}}
\end{figure}

The micro-well detector (MWD) is a type of gas proportional counter based on micro-patterned
electrodes \citep{dj2002a,dj2002b}.  Each sensing element consists of a
charge-amplifying well (Fig.~\ref{fig:well}).
The cathode and anode electrodes are deposited on opposite sides of an insulating substrate.
The well is formed as a cylindrical hole through the cathode and substrate, exposing the anode.
An array of such wells forms a detector, with the active tracking volume bounded by a drift
electrode.  Ionization electrons produced by the passage of
a fast charged particle drift toward the anodes and into the wells.  
An ionization avalanche occurs
in each well, where an intense electric field is set up by the voltage 
applied between the anode
and cathode.  Micro-well technology is very robust, and allows large
areas to be read out with good spatial ($\sim 100$ $\mu$m) and energy (18\% FWHM at 6 keV)
resolution at low cost.

We are working with our collaborators Thomas Jackson, Bo Bai, and Soyoun Jung 
at Penn State University to develop pixelized micro-well detectors (PMWDs) 
to enable true imaging of
charged particle tracks.  In this approach, each anode pad is connected to an element of a
thin-film transistor (TFT) array.  The individual transistor gates are connected in columns,
and the outputs are connected in
rows.  The gate drivers for each column are then activated sequentially, 
allowing the charge collected
on the anode pads to be read out by charge-integrating amplifiers at the 
end of each row.  Thus a
two-dimensional projected image of the charged particle track is recorded.  
The third dimension may
be determined by measuring the drift time of the ionization electrons using the
signals from the cathodes.
Ideally, the MWD and TFT array would be fabricated together as a single unit on a
robust, flexible substrate such as polyamide (e.g., Kapton\textsuperscript{TM}).

\section{Advanced Particle Tracker}
\label{sec:tracker}

\begin{figure}
\centering
\includegraphics[width=6.4cm]{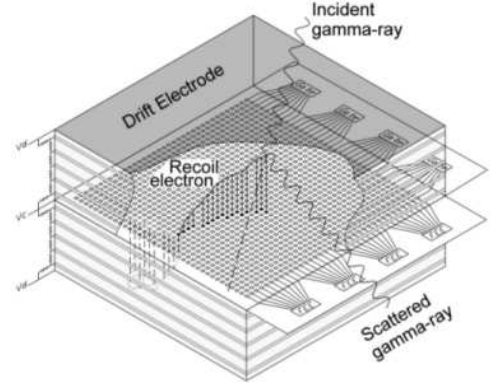}
\caption{TTID module consisting of two back-to-back PMWD-TFT arrays.  The track of
the recoil electron is imaged by the arrays, and the third dimension is calculated
from the charge drift time.\label{fig:ttid}}
\end{figure}

These combined PMWD-TFT arrays
will be assembled into modular detector units called three-dimensional 
track imaging detectors
(TTIDs), as shown in Fig.~\ref{fig:ttid}.  
Each TTID comprises two 30 cm $\times 30$ cm,
back-to-back PMWD-TFT arrays bounded
by drift electrodes (5 cm drift distance on each side) 
and field-shaping electrodes on the four walls.  The 
front end electronics, gate drivers, and timing electronics, together with their 
high-density interconnects, are distributed
around the periphery of the module and then folded up along the walls.

\begin{figure}
\centering
\includegraphics[width=4.4cm]{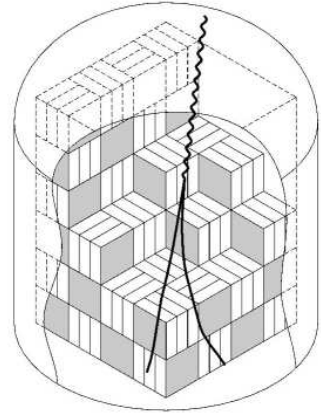}
\caption{Schematic of advanced particle tracker made up of TTID modules. 
\label{fig:tracker}}
\end{figure}

We have developed a concept for a large-volume charged particle tracker based on
TTID modules, shown schematically in Fig.~\ref{fig:tracker}.
We assume the PMWD-TFT arrays have a pitch of 200 $\mu$m and that a Xe/CO$_2$ gas
mixture (98\%/2\%) is used.  We also assume 10 ns timing resolution, which gives a
drift distance resolution of 140 $\mu$m for the maximum electron drift velocity in
this gas mixture.  The walls of the TTIDs are made of polyamide 300 $\mu$m 
thick.  The TTID modules are grouped into 30 cm $\times$ 30 cm cubes, and
arranged such that the drift direction in each cube is rotated 90$^{\circ}$ relative
to adjacent cubes; this gives a ``stereo'' view of extended tracks.  A 3 m diameter
pressure vessel would allow 61 cubes to be arranged inside per layer, 
giving 54900 cm$^2$ of 
geometric area while fitting within the payload fairing of a Delta III or IV launcher.

We have used Geant4.5.1 to simulate the performance of such a particle tracker.
Due to known errors in the multiple Coulomb scattering process in 
later versions of Geant4,
we used a slightly modified version of the multiple scattering class from version 3.2. 
In the following we describe our preliminary results for two 
possible uses of the charged particle tracker: an advanced Compton telescope and 
an advanced pair telescope.

\section{Advanced Compton Telescope}

The Advanced Compton Telescope (ACT) is envisioned as a $\sim 100$-fold increase in
sensitivity over that of CGRO/COMPTEL, the only Compton telescope that had enough
sensitivity to make astronomical observations \citep{schoen1993}.  Part of this
increase can be achieved by accepting larger Compton scatter angles, increasing
the effective area.  The rest will have to come from a dramatic decrease in the
telescope PSF, which reduces the area of the sky from which a given source's
photons could have originated.  This will reduce contamination both from internal
background and from nearby sources.

There are two components to the PSF of a Compton telescope \citep{bloserseeon}.  The first is
the error in the computed scatter angle
$\Delta\phi$.  (This is often referred to as the angular resolution measure, or ARM.)  This
width is determined by the spatial and energy resolution of the detectors that make up 
the telescope.  
The second component, $\Delta\theta$, is roughly given by the
error in the measurement of the recoil electron's initial direction, projected onto 
the plane perpendicular to the scattered photon direction.  COMPTEL was not able to
track the recoil electron at all, and so $\Delta\theta = 2\pi$.  The PSF for a single
photon was thus an annulus on the sky (the ``event circle'') with the 
diameter given by the scatter angle
$\phi$ and the width by $\Delta\phi$.  The total angular area of the PSF, 
$A = \sin\phi\Delta\phi\Delta\theta$, was therefore quite large for all but the smallest
scatter angles.  The ACT must accept scatter angles up to $\sim 120^{\circ}$ or greater,
and so good electron tracking may well be critical to keep the PSF, and therefore
background, within reasonable limits.  This is particularly true for good polarization
sensitivity, since the maximum polarization signal will be recorded for events with 
$\phi \sim 90^{\circ}$ \citep{bloserseeon}.

We have performed Geant4 simulations of an ACT concept using the gas particle tracker described
in Sec.~\ref{sec:tracker} with a depth of 2.7 m, filled with Xe/CO$_2$ gas with
a pressure of 3 atm.  The tracker is surrounded by a calorimeter made up of CsI pixels
to absorb the scattered photon.  We assumed CsI pixel dimensions of 5 mm $\times$ 5 mm, with
a depth of 5 cm on the sides of the tracker and 10 cm on the bottom.  We optimistically
assumed an energy resolution in the CsI of 5\% (FWHM) at 662 keV, which is one of the
main contributors to $\Delta\phi$.  In addition, we used the G4LECS package 
\citep{kippen2004} to calculate the Doppler broadening, a slight increase in $\Delta\phi$
due to scattering off bound electrons with unknown momenta in their atomic 
shells.  We simulated
2 MeV photons entering the telescope on-axis, and applied a simple detector response
(diffusion of drift electrons in the gas, energy resolution, and binning into pixels)
and event reconstruction to the output of the simulation.

\begin{figure}
\centering
\includegraphics[width=8cm]{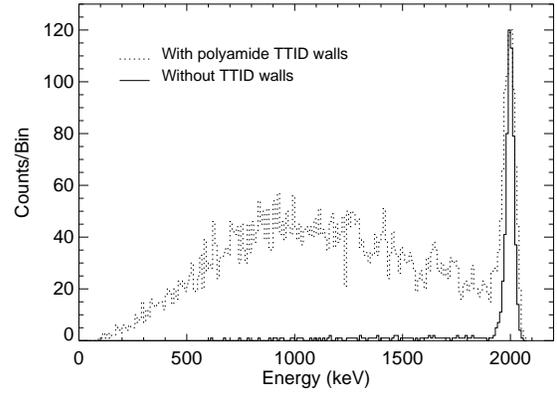}
\caption{Total energy spectrum recorded by ACT simulation.  Two cases are shown: 
300 $\mu$m polyamide TTID walls (dotted) and vacuum TTID walls (solid).  Spectra are
renormalized to have the same peak height.
\label{fig:spectrum}}
\end{figure}

Our results revealed a flaw in this ACT tracker design, which will be corrected in 
the next iteration: the recoil electrons lose energy in the polyamide walls of
the TTID modules, leading to large errors in the measured electron energy.  This can
be seen in Fig.~\ref{fig:spectrum}, which shows the total energy spectrum recorded
by the telescope for two cases: 300 $\mu$m thick polyamide TTID walls (dotted curve), and 
TTID walls artificially set to vacuum (solid curve).  
Such large energy loses in the tracker due to the TTID walls lead both to incorrect energy
measurements and to incorrect image reconstruction.  In future simulations, the TTID
modules will be made larger (50 cm on a side or more), and each will be surrounded by
its own mini-calorimeter, so that electrons leaving the tracking volume will immediately have
the remainder of their energy measured.  For the present scheme, the total energy
resolution at 2 MeV is about 5\% FWHM.

\begin{figure}
\centering
\includegraphics[width=9cm]{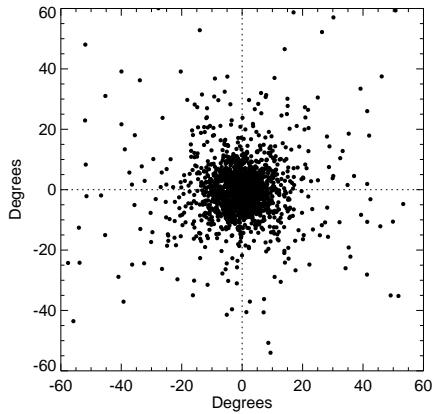}
\caption{ACT image at 2 MeV from simple back projection of individual events.  A fit with
a 2-dimensional Gaussian gives a $1\sigma$ width of $5.6^{\circ}$.
\label{fig:image}}
\end{figure}

To demonstrate the imaging capability of our ACT concept, we present in Fig.~\ref{fig:image}
the image derived at 2 MeV by simple back projection of individual events (without
polyamide walls).
A fit with a 2-dimensional Gaussian gives a $1\sigma$ width of $5.6^{\circ}$.  The 
ability to preform ``true'' imaging by back projection in this manner is a great 
simplification over COMPTEL and other instruments with poor electron tracking, which
require complicated reconstruction methods to produce an image from overlapping event
circles.  The effective area of this ACT concept at 2 MeV is $\sim 3000$ cm$^2$.  
Calculation of the sensitivity will require a realistic estimate of the in-flight 
background.

\section{Advanced Pair Telescope}

The angular resolution of a pair production telescope is limited by the multiple
scattering of the electron and positron in the detector material and by the unknown
recoil of the particle (nucleus or electron) in whose field the pair conversion took
place.  \citet{hunter2001} have shown that a pair telescope can nearly
achieve recoil-limited resolution, approaching 1 arcmin above a few GeV, if the
density of the tracking medium can be made less than $\sim 2 \times 10^{-5}$ 
radiation lengths (RL) per track measurement interval.  \citet{bloserspie} showed
that a pair telescope that achieved this density should in principle also be moderately
sensitive to polarization at $\sim 100$ MeV, since the azimuthal plane of the pair
is weakly correlated with the polarization vector.

We have simulated a concept for an Advanced Pair Telescope (APT) using the particle
tracker of Sec.~\ref{sec:tracker}.  The desired density of $2 \times 10^{-5}$ RL per
measurement is met with our 200 $\mu$m pitch if we use a Xe/CO$_2$ gas mixture at
a pressure of 1.5 atm.  A depth of 5.1 m then provides $\sim 0.5$ RL of total interaction
depth for pair conversion, similar to previous instruments.  The telescope does not
include a massive calorimeter; rather, the energy of the pair particles may be estimated by 
their average degree of scattering.  (For this reason the small amount of energy lost by 
electrons with several tens of MeV in the polyamide walls is of no concern for our APT
concept.)  The same detector response is applied as for the ACT simulations, and the 
incident photon direction is found by adding the momenta of the electron and positron.
We made use of a pair conversion class for Geant4 that includes the effects of polarization 
on the cross section \citep{bloserspie} to estimate the polarization sensitivity at 100 MeV.

\begin{figure}
\centering
\includegraphics[width=8cm]{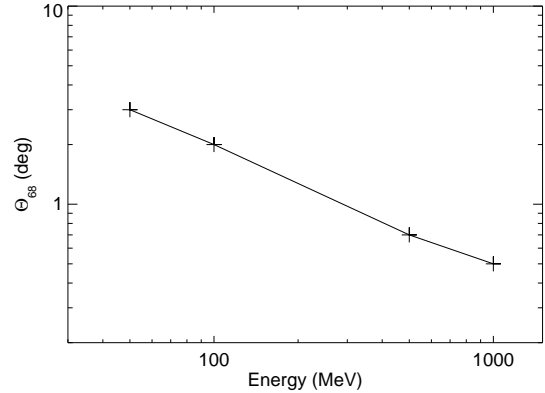}
\caption{Angular resolution of the APT concept as a function energy.  The resolution
is limited by diffusion of ionization electrons in the gas.
\label{fig:pair}}
\end{figure}

The derived angular resolution, defined as the angular radius $\Theta_{68}$ containing
68\% of the total events, is shown in Fig.~\ref{fig:pair}.
The resolution is nearly an order of magnitude worse than that predicted by 
\citet{hunter2001}.  This is due to the diffusion of ionization electrons as they
drift to the wells; for our Xe/CO$_2$ gas mixture 
the initial ionization cloud has spread by $\sigma_d \sim 1.1$ mm 
after drifting for 5 cm \citep{peisert}.
This blurs the particle tracks and makes them hard to measure precisely, especially
near the pair vertex where the tracks are close together.  The
current APT design does not have a useful polarization sensitivity for the same
reason.  We are currently investigating means of reducing the diffusion in order to
achieve the desired angular resolution and polarization sensitivity.  The most promising
method appears to be the addition of an electronegative gas (e.g. CS$_2$) which causes
the ionization electrons to attach themselves to the ions, which then drift with 
much smaller diffusion \citep{martoff}.

\section*{Acknowledgments}

We thank G. Depaola and F. Longo for the Geant4 pair polarization class and 
T. Burnett for the modified Geant4.3.2 multiple scattering class.
This work was performed while the author held a National Research Council Research
Associateship Award at NASA/GSFC.

% The following bibliography was produced with
%   \bibliographystyle{aa}
%   \bibliography{esapub}
% The results are inserted directly here to simplify
% the demonstration.

\end{document}